\documentclass[twocolumn]{aa}
\usepackage{graphicx}
\usepackage[figuresright]{rotating}
\usepackage{txfonts}
\usepackage{booktabs}
\begin{document}
\title{{A Search for Near-Infrared Molecular Hydrogen Emission in the CTTS LkH$\alpha$ 264 and the 
debris disk 49 Cet}
\thanks{Based on observations collected at the European Southern
Observatory, Chile (program ID 60.A-9064(A)).}}
\titlerunning{NIR H$_2$ emission from protoplanetary disks}
   \author{A. Carmona
          \inst{1,2}
          \and
          M.E van den Ancker \inst{2}
          \and
          Th. Henning \inst{1}
          \and
          M. Goto \inst{1}
          \and
          D. Fedele\inst{2,3}
          \and
          B.Stecklum \inst{4}
          }
   \offprints{A. Carmona\\
              \email{carmona@mpia.de}}             
   \institute{
              Max Planck Institute for Astronomy, K\"onigstuhl 17, 69117 Heidelberg, Germany
           \and  
              European Southern Observatory, 
              Karl Schwarzschild Strasse 2 , 85748 Garching bei M\"unchen, Germany
           \and
              Dipartimento di Astronomia, Universit\`a di Padova, 
              Vicolo dell'Osservatorio 2, 35122 Padova, Italy  
           \and
           	  Th\"uringer Landessternwarte Tautenburg, 
	  		  Sternwarte 5, 07778 Tautenburg, Germany	             
           }
   \date{Accepted by A\&A 9 October 2007}
   \abstract{
    We report on the first results of a search for molecular hydrogen emission from 
    protoplanetary disks using CRIRES, ESO's new VLT Adaptive Optics 
    high resolution near-infrared spectrograph.
    We observed the classical T Tauri star  LkH$\alpha$ 264 and the debris disk 49 Cet,
    and searched for $\upsilon= 1-0$ S(1) H$_2$ emission at 2.1218 $\mu$m,
    $\upsilon = 1-0$ S(0) H$_2$ emission at 2.2233 $\mu$m and 
    $\upsilon = 2-1$ S(1) H$_2$ emission at 2.2477 $\mu$m. 
    The H$_2$ line at 2.1218 $\mu$m is detected in LkH$\alpha$ 264 confirming the
    previous observations by Itoh et al. (2003).
    In addition,
    our CRIRES spectra reveal the previously observed but not detected 
    H$_2$ line at 2.2233 $\mu$m in LkH$\alpha$ 264.
    An upper limit of 5.3 $\times 10^{-16}$ ergs s$^{-1}$ cm$^{-2}$ 
    on the $\upsilon = 2-1$ S(1) H$_2$ line flux in LkH$\alpha$ 264 is derived.    
    The detected lines coincide with the rest velocity of LkH$\alpha$ 264.
    They have a FWHM of $\sim$20 km s$^{-1}$. 
    This is strongly suggestive of a disk origin for the lines.
    These observations are the first simultaneous detection of 
    $\upsilon = 1-0$ S(1) and $\upsilon = 1-0$ S(0) H$_2$ emission from a protoplanetary disk.    
    49 Cet does not exhibit H$_2$ emission in any of the three observed lines.
    We derive the mass of optically thin 
    H$_2$ at $T\sim1500$ K  in the inner disk
    of LkH$\alpha$ 264 and derive stringent limits in the case of 49 Cet at the same temperature.
    There are a few lunar masses of optically thin hot H$_2$ in the inner disk ($\sim$ 0.1 AU) 
    of LkH$\alpha$ 264,
    and less than a tenth of a lunar mass of hot H$_2$ in the inner disk of 49 Cet.
    The measured 1-0 S(0)/1-0 S(1) and 
    2-1 S(1)/1-0 S(1) line ratios in LkH$\alpha$ 264 indicate 
    that the H$_2$ emitting gas 
    is at a temperature lower than 1500 K and that the H$_2$ is most likely thermally
    excited by UV photons. 
    The $\upsilon = 1-0$ S(1) H$_2$ line in LkH$\alpha$ 264 is single peaked
    and spatially unresolved.
    Modeling of the shape of the line suggests that the disk  
    should be seen close to face-on ($i<35^{\rm\,o}$) and that the
    line is emitted within a few AU of the LkH$\alpha$ 264 disk. 
    A comparative analysis of the physical properties of classical T Tauri stars
    in which the H$_2$ $\upsilon = 1-0$ S(1) line has been detected and non-detected indicates
    that the presence of H$_2$ emission is correlated with the magnitude of
    the UV excess and the strength of the H$\alpha$ line.      
    The lack of H$_2$ emission in the NIR spectra of 49 Cet
    and the absence of H$\alpha$ emission suggest that the gas in the 
    inner disk of 49 Cet has dissipated.
    These results combined with previous detections of $^{12}$CO emission at sub-mm wavelengths 
    indicate that the disk surrounding  49 Cet should have an inner hole. 
    We favor inner disk dissipation by inside-out photoevaporation, or
    the presence of an unseen low-mass companion as the most likely explanations
    for the lack of gas in the inner disk of 49 Cet.

   }
   \maketitle
%
%
%
\begin{table*}
\caption{Stellar physical properties.}             
\label{table:1}      
\centering                          
\begin{tabular}{l r c c c c c c c c c c c c c c c }        
\hline\hline                 
Star          & Sp.T.       & $T_{eff}$ & $d$   & 
Age           & CO sub-mm   & $M_{disk}^{\dag}$     & object \\
              &             & [K]       &  [pc] & 
[Myr]         &             & [M$_J$]        &         
& \\    
\hline
\addlinespace[5pt]                                   
LkH$\alpha$ 264 & K5Ve$^a$       & 4350$^b$      & 300$^c$ &
2$^d$               & ...                        & 85$^a$  & CTTS         \\
49 Cet          & A1V$^e$	 & 9970$^b$          & 61$^f$      &
20$^g$          &$^{12}{\rm CO}\,J=3-2^h~J=2-1^i$       & 0.4$\,^j$ & debris disk         \\
\hline
\addlinespace[5pt]
\multicolumn{9}{l}{$^a$ Itoh et al. (2003b).
\,$^b$ Kenyon \& Hartmann (1995). 
\,$^c$ Strai{\v z}ys et al. (2002).
\,$^d$ Jayawardhana et al. (2001).}\\
\multicolumn{9}{l}{
$^e$ From Chen et al. (2006).
\,$^f$ Hipparcos catalogue.
\,$^g$ Zuckerman \& Song (2004).
\,$^h$ Dent et al. (2005)}\\
\multicolumn{9}{l}{$^i$ Zuckerman et al. (1995).
\,$^j$ Thi et al. (2001) and Bockl\'ee-Morvan et al. (1995). }\\
\multicolumn{9}{l}{$^{\dag} M_{disk}$ refers to the total mass in the disk deduced from 
mm dust continuum emission assuming a }\\
\multicolumn{9}{l}{gas to dust ratio of 100.}\\
\end{tabular}
\end{table*}

\section{Introduction}
The discovery of extrasolar planets triggered  an
increasing interest in the physical mechanisms behind the process of planet formation.
Many recent efforts have been directed to the study of 
disks surrounding pre-main-sequence stars.
Observational and theoretical evidence suggests  
that planets are forming in these disks.
The observational characterization of the physical structure and dynamics of
the gas and dust in protoplanetary disks
is of paramount importance for understanding the process of planet formation.

In the inner 1 AU of protoplanetary disks,
intense UV or X-ray heating can bring the gas temperatures to 
a few thousand Kelvin.
At these high temperatures,
ro-vibrational transitions of H$_2$ are excited and 
a rich spectrum of H$_2$ lines in the near-infrared is expected to be produced.
The study of H$_2$ quiescent ro-vibrational emission\footnote{By quiescent emission we mean
emission at the rest velocity of the star.
H$_2$ emission can also be produced by shocked gas associated with outflows.
However, in such a case the emission is expected to be doppler shifted more than 20 km s$^{-1}$
with respect to the rest velocity of the star.}
towards pre-main-sequence stars with disks 
offers the opportunity to address the question of the presence of hot gas in the disk,
by probing the temperature and density in the innermost regions where terrestrial planets
are expected to form.
For example, the H$_2$ $\upsilon =1-0$ S(1) line at 2.1218 $\mu$m (one of the strongest H$_2$ ro-vibrational lines) 
is sensitive to a few lunar masses of gas. Therefore, the absence of the line would be
strongly suggestive of little or no hot gas in the systems.

In this paper we present the first results of a sensitive search for near-infrared H$_2$  
emission from protoplanetary disks using CRIRES, 
ESO's new VLT near-infrared high-resolution spectrograph. 
We searched for  
the H$_2$ $\upsilon =1-0$ S(1) line at  2.1218 $\mu$m,
H$_2$ $\upsilon = 1-0$ S(0) line at 2.2233 $\mu$m
and H$_2$ $\upsilon = 2-1$ S(1) line at 2.2477 $\mu$m,  
towards LkH$\alpha$ 264, a classical T Tauri star with previously 
reported detections of the $\upsilon =1-0$ S(1)
line by Itoh et al. (2003), 
and 49 Cet, 
a debris disk with evidence of a large reservoir 
of cold gas at sub-mm wavelengths (Dent et al. 2005, Zuckerman et al. 1995).
For a summary of the stellar physical properties of LkH$\alpha$ 264 and 49 Cet see Table 1.
We confirm the detection of H$_2$ emission at 2.1218 $\mu$m in LkH$\alpha$ 264,
and announce, for the first time, the detection of H$_2$ emission at 2.2233 $\mu$m  from a disk (LkH$\alpha$ 264).
We report the non-detections of H$_2$ emission at 2.2477 $\mu$m in LkH$\alpha$ 264,
and at 2.1218, 2.2233 and 2.2477 $\mu$m in 49 Cet.

The paper is organized as follows.
We start with a description of the observations and the data reduction.
In section 3 we present our results and calculate the mass limits for 
the hot ($T\sim1500$ K) H$_2$ in the systems. 
In Section 4, based upon the measured 1-0 S(0)/1-0 S(1) and 
2-1 S(1)/1-0 S(1) line ratios in LkH$\alpha$ 264, 
we determine the excitation mechanism of 
the observed H$_2$ emission.
By modeling of the shape of the $\upsilon = 1-0$ S(0) H$_2$ line,
we derive constraints on the H$_2$ emitting region and 
the inclination of the disk around LkH$\alpha$ 264.
Finally we discuss the disk properties of the stars in which H$_2$
emission has been detected and the prospects for future investigations.
Our conclusions are presented in Section 5. 

\begin{table*}
\caption{Summary of the observations.}             
\label{table:observations}      
\centering                          
\begin{tabular}{@{}l c l c c c c c l c c c c c c c @{}}        
\hline\hline                 
Star            & $\lambda$      & Date      & UT      & 
$t_{exp}$       & Airmass        & seeing    &  
Calibrator $^a$  & $t_{exp}$      & Airmass   & seeing  \\
                & [$\mu$m]       &           & [hh:mm]  & 
[s]             &                & [arcsec]  &             
                & [s]            &           & [arcsec] \\
\hline                        
LkH$\alpha$ 264 & 2.1218      &  8 Nov 2006  & 06:25   &
720             & 1.4            & 1.2       &            
HIP 13327    & 160            & 1.3       & 0.9 \\
                & 2.2233, 2.2477       &  8 Nov 2006  & 06:52   &
720             & 1.4            & 1.0       &     
HIP 13327    & 160            & 1.3       & 0.9 \\ 
49 Cet          & 2.1218         &  9 Nov 2006  & 02:38   &
240             & 1.1            & 0.8       &  
HIP 8497    & 40             & 1.0       & 1.2          \\
                & 2.2233, 2.2477       &  9 Nov 2006  & 03:07   &
240             & 1.2            & 0.7       &  
HIP 8497     & 40             & 1.0       & 1.2          \\
\hline  
\addlinespace[5pt]
\multicolumn{10}{l}{$^a$ Spectrophotometric standard stars were observed
immediately following the science observations.} 
\end{tabular}
\end{table*}
\begin{table*}
\caption{H$_2$ line fluxes and upper limits measured }             
\label{table:results}      
\centering                          
\begin{tabular}{@{} l l c c c c c c c c @{} }        
\hline\hline
\addlinespace[5pt]
           & & \multicolumn{3}{c}{LkH$\alpha$ 264}
           & \multicolumn{3}{c}{49 Cet }\\
\cmidrule(r){3-5}\cmidrule(r){6-8}          
            & &1-0 S(1) & 1-0 S(0) & 2-1 S(1) &
               1-0 S(1) & 1-0 S(0) & 2-1 S(1) \\
            & & 2.1218 $\mu$m & 2.2233 $\mu$m      & 2.2477 $\mu$m 
              & 2.1218 $\mu$m & 2.2233 $\mu$m      & 2.2477 $\mu$m \\
\hline 
\addlinespace[5pt]          
 Continuum  & [ergs s$^{-1}$ cm$^{-2}~\mu$m$^{-1}$] 
            & $9.2\times 10^{-11}$   &$\,\,1.5\times 10^{-10}$ &$\,\,\,1.6\times 10^{-10}$
            & $\,3.5\times 10^{-9}$  &$\,1.7\times 10^{-9}$    &$\,\,1.5\times 10^{-9}$\\
 3$\sigma$  & [ergs s$^{-1}$ cm$^{-2}~\mu$m$^{-1}$] 
 			& $4.2\times 10^{-12}$   &$\,\,8.2\times 10^{-12}$ &$\,\,\,1.0\times 10^{-11}$
			& $\,1.0\times 10^{-10}$ &$\,\,1.8\times 10^{-10}$ &$\,\,\,3.1\times 10^{-10}$\\
 Integrated Line Flux$^{a}$ & [ergs s$^{-1}$ cm$^{-2}$] 
 			& $3.0\times 10^{-15}$   & $1.0\times 10^{-15}$    & $<5.3\times 10^{-16}$  
			&$<5.4\times 10^{-15}$   & $<8.9\times 10^{-15}$   & $<1.6\times 10^{-14}$ \\
\hline                                   
\addlinespace[5pt]
\multicolumn{8}{l}{$^a$ For the calculation of upper limits, we assumed that the FWHM of the line is 6.6 km s$^{-1}$. 
}\\
\end{tabular}
\end{table*}
\section{Observations}
We obtained high-resolution ($R\sim45000$)\footnote{The spectral resolving power of our CRIRES observations
was determined by the FWHM of the Gaussian fit of an unresolved skyline. 
A $\lambda/\Delta \lambda \sim45000$ at 2.12 $\mu$m corresponds to a resolution of $\sim$6.6 km s$^{-1}$.
This FWHM is sampled on 5 pixels of the CRIRES detector. }
near-infrared spectra of LKH$\alpha$ 264 and 49 Cet,
on 2006 November 8-9,  
using the ESO-VLT cryogenic high-resolution infrared echelle spectrograph CRIRES (K\"{a}ufl et al. 2004),
mounted on ESO UT1 "Antu" 8-m telescope atop Cerro Paranal Chile,
during the CRIRES science-verification phase. 
CRIRES uses a mosaic of four Aladdin III InSb arrays providing an effective 4096 x 512
detector array in the focal plane. 
Adaptive Optics (MACAO - Multi-Applications Curvature Adaptive Optics) 
was used to optimize the signal-to-noise ratio and the spatial resolution.
The science targets were used as natural guide stars.

Our observations were performed using a 31" long, 0.4" wide, 
north-south oriented slit.
The observations were made by nodding the telescope 10" along the slit.
To correct for bad pixels and decrease systematics due to the
detector, a random jitter smaller than 2" was added to 
the telescope in addition to the nodding offset at each nodding position. 
For the telluric correction spectrophotometric standard stars at similar airmass to the science target
were observed immediately following the science observations. 

The observations were performed employing the wave-ID 27/1/n and the wave-ID 25/-1/n,
providing a spectral coverage from 2.0871 to 2.1339 $\mu$m and from 2.2002 to 2.2552 $\mu$m respectively.
To obtain the wavelength calibration, 
observations of an internal Th-Ar calibration lamp with 3x30 second exposures
were executed immediately following the target and standard star spectroscopy observations at each wave-ID setting.
A summary of the observations is provided in Table 2.

\subsection{Data Reduction}
The data was reduced using the CRIRES pipeline and the ESO/CPL recipes.
In each chip, 
raw image frames at each nodding position 
were flat-field corrected, 
then image pairs in the nodding sequence (AB) were subtracted and
averaged resulting in a combined frame ($F_{combined}=(A-B)/2$), 
thereby correcting for the sky-background.
The frames at each nodding position were corrected 
for jittering using the jitter information from the fits headers.
The ensemble of combined frames were stacked in one single 2D frame.
The spectrum was extracted by summing the number of counts inside the PSF in the 
dispersion direction in the 2D spectrum.
The absolute wavelength calibration was obtained by cross-correlation with the Th-Ar lamp frame taken 
immediately after the science exposure.
The wavelength calibration was done for each chip independently.
The one-dimensional spectrum was divided by the exposure time   
$t_{exp}$ (see Table 2).
 
To correct for telluric absorption, 
the one-dimensional extracted science spectrum was divided by   
the one-dimensional extracted spectrum of the standard star. 
The standard star spectrum  was corrected 
for differences in air-mass and air-pressure with respect to the science target spectrum
employing the method described by Carmona et al. (2005). 
Small offsets of a fraction of a pixel in the wavelength direction were applied
to the standard star spectrum
until the best signal-to-noise in the corrected science spectra was obtained.  

Absolute flux calibration was found
by multiplying the telluric corrected spectrum by the flux of the standard star at the
wavelengths observed. The flux of the standard star in the K band was found from the $K$ magnitude
of the standard star using Vega as the reference star\footnote{The flux of Vega in the K band used is
4.14 $\times 10^{-10}$ W m$^{-2}$ $\mu$m$^{-1}$ (Cox, 2000).}.
The absolute flux calibration is accurate at the 20\% level. 
Imperfections in the telluric correction are the principal source of
uncertainty.

\begin{figure*}
\centering
\includegraphics[angle=0,width=\textwidth]{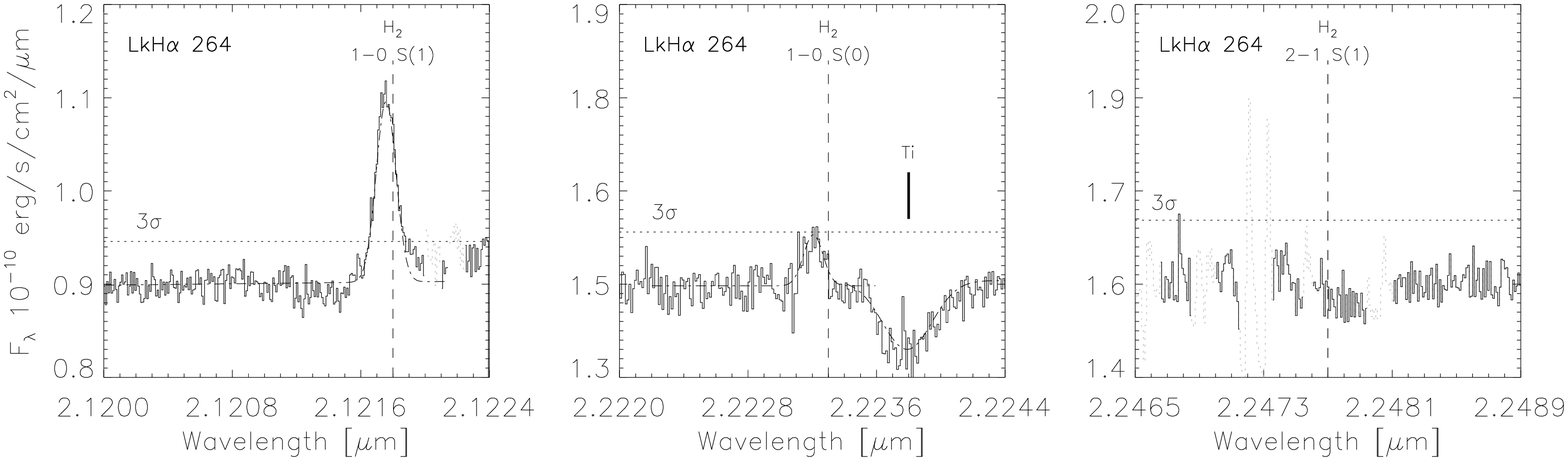}
\includegraphics[angle=0,width=\textwidth]{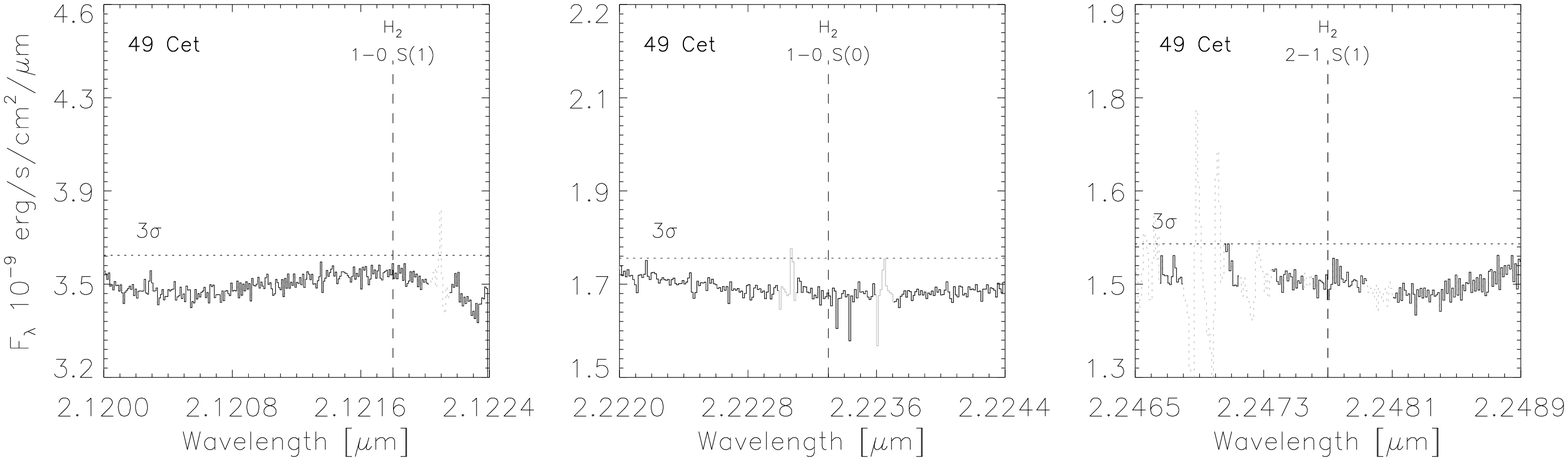}
    \caption{CRIRES spectra of LkH$\alpha$ 264 (upper panels) and 49 Cet (lower panels) in the
      regions of the H$_2$ $\upsilon$=1-0 S(1), H$_2$ $\upsilon$=1-0 S(0) and H$_2$ $\upsilon$=2-1 S(1)
      emission lines. The H$_2$ $\upsilon$=1-0 S(1)
      and the H$_2$ $\upsilon$=1-0 S(0) lines are detected in LkH$\alpha$ 264. 
      Photospheric Ti features at 2.2217 (not shown) and 2.2238 $\mu$m 
      (central upper panel) are observed in LkH$\alpha$ 264.
      The Gaussian fits to the detected lines are illustrated in dash-dot lines.
      The H$_2$ $\upsilon$=2-1 S(1) line is not present in LkH$\alpha$ 264.
      In the case of 49 Cet none of the three H$_2$ features are present in emission or absorption.
      Horizontal dotted lines show the 3$\sigma$ continuum flux limits.
      The spectra are not corrected for ${\rm V}_{\rm LSR}$ of the star. 
      Regions of poor telluric correction are in gray-dotted lines in the spectra.
               }
\end{figure*}

\section{Results}
We present in Figure 1 the spectra of LkH$\alpha$ 264 and 49 Cet.
LkH$\alpha$ 264 exhibits H$_2~\upsilon = 1-0$ S(1)  emission at 2.1218 $\mu$m
and the H$_2~\upsilon = 1-0$ S(0) feature at 2.2233 $\mu$m.
Our observations confirm the previous detections of the H$_2~\upsilon = 1-0$ S(1)  line reported
by Itoh et al. (2003). 
The line flux measured from our CRIRES spectrum is $\sim$ 50\% fainter than the line reported by Itoh et al. (2003) (see Section 3.2).
In contrast to Itoh et al. (2003) the H$_2~\upsilon = 1-0$ S(0) line
is detected in our CRIRES spectra of LkH$\alpha$ 264.
The H$_2~\upsilon = 2-1$ S(1) line is not seen in LkH$\alpha$ 264.
The Si line at 2.1210 $\mu$m 
reported by Itoh et al. (2003) is not confirmed by our CRIRES spectrum.
We observe Ti absorption lines at 2.2217 and 2.2238 $\mu$m 
of FWHM of 48.9 km s$^{-1}$ and EW 0.2 \AA. These lines are
gravity sensitive, and would suggest that the underlying photosphere is of a
late K dwarf (see, e.g., Greene \& Lada 2002) in agreement with the spectral type
K5Ve of LkH$\alpha$ 264 .  
The broadening of the lines indicates a ${\rm v} ~{\rm sin\,}i$ of 40 km s$^{-1}$ in LkH$\alpha$ 264.  
In the case of 49 Cet none of the three H$_2$ features are present in emission or absorption.
The spectrum does not exhibit photospheric absorption features.
We summarize our results in Table 3.
We should note that the 49 Cet observations were less deep than the LkH$\alpha$ 264,
however, the distance of the target (61 pc) compensate for this.

\subsection{Upper Flux Limits on H$_2$ Emission in 49 Cet}
3$\sigma$ upper limits for the integrated line flux of the 
observed H$_2$ lines in 49 Cet
were determined 
by calculating the standard deviation of the continuum flux 
in the vicinity of the H$_2$ features 
and multiplying 3 times the standard deviation (see horizontal dotted lines in Figure 1) 
times the FWHM of a CRIRES unresolved line (4.95 $\times 10^{-5} \mu$m). 
They are reported in Table 3.
The continuum is spatially unresolved.
The mean PSF FWHM measured in the continuum is 3.7 and 6.0 pixels 
for the spectrum at 2.12 and 2.22 $\mu$m respectively. 
Using the CRIRES pixel scale of 0.086 arcsec/pixel,  
we obtain a PSF FWHM of 0.3" and 0.5" for each spectrum.
At the distance of 49 Cet (61 pc), that corresponds to 20 and 31 AU respectively.
We conclude that the size of the NIR continuum-emitting region of 49 Cet has 
an upper limit of 20 AU. 

\subsection{Molecular Hydrogen Emission in LkH$\alpha$ 264}
The central wavelength of the $\upsilon = 1-0$ S(1) H$_2$ emission in LkH$\alpha$ 264
was measured to be 2.121757 $\pm$ 0.000005 $\mu$m by a Gaussian fit. 
Assuming an error on the wavelength calibration of $\sim$1.0 km s$^{-1}$ (1 pixel),
it corresponds to a velocity shift of -6.0 $\pm$ 1.0 km s$^{-1}$.
At the time of the observations the velocity correction\footnote{We employed the rvcorrect function of IRAF to calculate the velocity shift.} due to the motion of the Earth was +0.4 km s$^{-1}$.
The velocity shift is therefore -5.6 $\pm$ 1.0 km s$^{-1}$,
in agreement with the center of the line of -5.1 $\pm$ 1.2 km s$^{-1}$ by Itoh et al. (2003) 
and the rest velocity of the star of -5.9 $\pm$ 1.2 km s$^{-1}$ by Itoh et al. (2003) 
and -4.2$\pm$ 2.5 km s$^{-1}$ by Hearty et al. (2000).
Our observations confirm that the H$_2$ emission observed is coincident with the rest velocity of the star. 
The detected H$_2$ $\upsilon = 1-0$ S(1) emission line is symmetric.
The observed FWHM of the line\footnote{Convolved with the instrumental width of 6.6 km s$^{-1}$.} 
is 20.6 $\pm$ 1 km s$^{-1}$.
The Equivalent Width (EW) of the line is -0.32 $\pm$ 0.01 \AA,
and the integrated line flux is  3.0 $\times 10^{-15}$ erg s$^{-1}$ cm$^{-2}$.
The observed line is 10 km s$^{-1}$ narrower and slightly fainter than the line observed by
Itoh et al. (2003).
The line flux observed is within the range of 1 to 7 $\times 10^{-15}$ erg s$^{-1}$ cm$^{-2}$
line fluxes reported towards other classical T Tauri stars (Weintraub et al. 2000, Bary et al. 2003). 

The  H$_2$ $\upsilon = 1-0$ S(0) feature at 2.2233 $\mu$m is detected with a 3$\sigma$ level
confidence (see Figure 1). 
Employing a Gaussian fit, 
the central wavelength of the line found is 2.22321 $\pm$ 0.00005 $\mu$m.
This corresponds to a velocity shift of -12 $\pm$ 7 km s$^{-1}$ which is 
in agreement with the velocity shift found in the $\upsilon = 1-0$ S(1) line.
The error in the determination of the center of the $\upsilon = 1-0$ S(0) line
is larger because the line is detected with a much smaller confidence level.
The FWHM of the line is 19.8 $\pm$ 1 km s$^{-1}$.
The EW of the line is -0.07 $\pm$ 0.01 \AA,
and the integrated line flux is  1.0 $\times 10^{-15}$ erg s$^{-1}$ cm$^{-2}$.
This line flux is smaller than the previous flux upper limits by Itoh et al. (2003)
demonstrating the improvement in sensitivity reached by CRIRES.

The H$_2$  $\upsilon = 1-0$ S(1) emission is spatially unresolved.
The mean PSF FWHM in the continuum measured is $\approx$4.2 pixels.
Using the CRIRES pixel scale of 0.086 arcsec/pixel and 
a distance of 300 AU for LkH$\alpha$ 264,
we obtain a PSF FWHM of $\approx$0.36" indicating 
that the $\upsilon = 1-0$ S(1) line is produced  
in the inner 50 AU of the LkH$\alpha$ 264 disk.
The H$_2$  $\upsilon = 1-0$ S(0) emission is also spatially unresolved.
The mean PSF FWHM in the continuum measured is $\approx$ 0.58" (6.8 pixels)
corresponding to an upper limit of 90 AU for the H$_2$ $\upsilon = 1-0$ S(0) 
emitting region.
The very similar FWHM of the H$_2~\upsilon = 1-0$ S(0) and the 
H$_2~\upsilon = 1-0$ S(1) ($\sim$ 20 km s$^{-1}$)
suggests that the gas responsible for the H$_2$ emission is located in similar 
regions of LkH$\alpha$ 264. 
Since both H$_2$ lines are spatially unresolved,
and both lines presumably come from the same region,
we conclude that the H$_2$ emitting region should be in the inner
50 AU of the LkH$\alpha$ 264 disk. In Section 4.2, from the line ratio
of the detected lines (i.e. temperature of the H$_2$ emitting the lines) 
and the shape of the line profile, we set more stringent
limits to the location of the emission.

Employing a similar approach as described for 49 Cet,
we derived an upper limit for the flux of 5.3 $\times 10^{-16}$ erg s$^{-1}$ cm$^{-2}$
for the H$_2$ $\upsilon = 2-1$ S(1) feature at 2.2477 $\mu$m in LkH$\alpha$ 264.
Assuming an error of 20\% 
in the flux calibration of the spectra,
the 1-0 S(0)/1-0 S(1) line ratio is 0.33 $\pm$ 0.1 
and 
the 2-1 S(1)/1-0 S(1) line ratio is $<$0.2.
These line ratios are consistent with the line ratios of a gas at LTE at a temperature 
cooler than 1500 K (Mouri 1994).

\subsection{Mass of Hot H$_2$ in LkH$\alpha$ 264 and 49 Cet}
Assuming optically thin emission and a source size equal or smaller to the beam size 
of the telescope, 
the mass of hot H$_2$ gas in $M_{\oplus}$ was determined from the 
$\upsilon = 1-0$ S(1) line flux employing (Bary et al. 2003, Thi et al. 2001)
\begin{equation}
M({\rm H}_2)_{\,\upsilon =1-0\,S(1)}= 5.84 \times 10^{-15}\frac{4\pi F_{ul}D^{2}}{E_{ul}A_{ul}~\chi_{\upsilon,J}(T)}
\end{equation}
with $F_{ul}$ being the $\upsilon = 1-0$ S(1) line flux or the flux upper limit, 
$D$ the distance in pc to the source,
$E_{ul}$ the energy difference in ergs between the levels $u$ and $l$ of the 
transition (9.3338$\times10^{-13}$ ergs), $A_{ul}$
the Einstein coefficient ($A_{10}=2.09 \times 10^{-7}$ s$^{-1}$) and  
$\chi_{\upsilon,J}(T)$
the level populations at temperature $T$ of the H$_2$ gas at the upper level $\upsilon$,\,J of the transition
(Bary et al. 2003).
Under LTE conditions at 1500 K, $\chi_{\upsilon,J}(T)=5.44\times10^{-3}$. 
In LkH$_\alpha$ 264 the mass of hot gas is $\approx 0.019 M_{\oplus}$ ($\sim$1.5 M$_{\rm Moon}$).
Since the flux observed by our CRIRES observations is 50\% lower to that
reported by Itoh et al. (2003) the derived mass is $\approx$ 50\% lower.
Using the same set of equations,  the upper limit to the mass of H$_2$ at $T=$1500 K 
obtained for 49 Cet is $0.0014 M_{\oplus}$ ($\sim$0.1 M$_{\rm Moon}$).
Note that if a lower temperature is assumed, the level populations are smaller 
and the deduced mass of H$_2$ is larger. 

Comparing the disk masses deduced from observations of dust continuum
at mm wavelengths (see Table 1) with the gas mass probed by the $\upsilon = 1-0$ S(1) H$_2$
line (see Table 3), 
we can observe that the amount of gas that is probed by the $\upsilon = 1-0$ S(1) H$_2$
line is very small with respect to the total amount of gas inferred to be in the disk.
Bary et al. (2003) suggest that a conversion factor of $10^{7}-10^{9}$ could be used 
for deducing the total mass of the gas from the masses obtained from the $\upsilon = 1-0$ S(1) H$_2$ line.
Applying such a conversion factor
for LkH$\alpha$ 264 we obtain a total disk mass of 0.5 to 50 M$_{\odot}$ and
for 49 Cet a total disk mass of $<$ 0.04 to 0.4 M$_{\odot}$.
In the case of 49 Cet the deduced upper limits to the disk mass are in 
agreement with the mass deduced from mm dust continuum observations.
In the case of LkH$\alpha$ 264 the total mass deduced 
is much too high to be consistent with the mass obtained from 
observations at mm wavelengths. In addition, the estimate is
unrealistic since a disk this massive 
   would have fragmented under the influence of its own gravity.
  
\section{Discussion}
\subsection{The excitation mechanism of the H$_2$ line in LkH$\alpha$ 264}
H$_2$ emission can be the result
of thermal (collisions)  and
non-thermal (radiative decay from excited electronic states)
excitation mechanisms.
In the thermal case, 
the gas  is heated either by shocks, X-rays or UV-photons.
In this case, the H$_2$ spectrum is characterized by a single excitation temperature
typically between 1000 and 2000 K.  
In the non-thermal case, the electronic excitation results from the
absorption of a UV photon in the Lyman-Werner band (912-1108 \AA)
or the collisions with a fast electron  due to X-ray ionization (Mouri 1994).

The first step in our analysis is to determine whether the H$_2$ emission observed in
LkH$\alpha$ 264 originates in an outflow (shock excited emission) or in a disk. 
The small velocity shift, the line shape (well reproduced by a disk model, see \S 4.2),
and the fact that the emission is spatially unresolved   
are not in favor of shock excited H$_2$.
An additional strong argument against shock excitation of H$_2$ 
is that  LkHa 264 does not exhibit [OI] forbidden emission at 6300 \AA~(Cohen and Kuhi 1979);
a classical signature of outflows in T Tauri stars.
The lack of this line indicates that in LkH$\alpha$ 264 the outflow is not present or at least that 
it is very weak. We conclude that the H$_2$ emission observed in LkH$\alpha$ 264
originates very likely in a disk.  

The thermal and non-thermal excitation mechanisms are distinguishable on the basis of 
line ratios (Mouri 1994 and references there in).
With Figure 3b of Mouri (1994),
we find that the measured 1-0 S(0)/1-0 S(1) (0.33 $\pm$ 0.1)
and the 2-1 S(1)/1-0 S(1) ($<$0.2) line ratios
in LkH$\alpha$ 264 are consistent with
thermal emission of a gas cooler than 1500 K.
If the distribution of errors is assumed Gaussian,
then 
a 3$\sigma$ error of 0.1 in the 1-0 S(0)/1-0 S(1) ratio implies
that there is a 90\% probability that 1-0 S(0)/1-0 S(1) $>$ 0.28.
Therefore, the most likely scenario 
is that the H$_2$ emitting gas 
is at a temperature cooler than 1000 K and that the H$_2$ is thermally
excited by UV photons\footnote{Itoh et al. (2003) employing upper limits 
to the 1-0 S(0) emission suggested that the 1-0 S(0)/1-0 S(1) is smaller than 0.26
for LkH$\alpha$ 264. We detect the 1-0 S(0) line and find that most 
likely this line ratio is higher than 0.28.}.  
LkH$\alpha$ 264 is also an X-ray source (Hearty et al. 2000).
Nevertheless, given the line 1-0 S(0)/1-0 S(1) ratio measured,
the probability that the heating mechanism is X-ray excitation is less than 
1\% (1-0 S(0)/1-0 S(1) $<$ 0.23). 
The conclusion that the H$_2$ observed emission is very likely due to UV-photons
is supported by the fact that LkH$\alpha$ 264 has a strong UV excess ($U-V=0.37$,
Bastian \& Mundt 1979). 
 
\subsection{H$_2$ Emitting Region and Inclination of the Disk around LkH$\alpha$ 264}
The spectral resolution of CRIRES ($\approx$ 6.6 km s$^{-1}$) 
and the thermal width of a 1500 K line ($\approx$ 2.4 km s$^{-1}$) 
are significantly smaller than the FWHM of 20 km s$^{-1}$ 
of the H$_2$ lines observed in LkH$\alpha$ 264.
Therefore, the line width must be linked to the dynamics of the gas in
the region that is emitting the line.
If the molecular hydrogen emission in LkH$\alpha$ 264 originates in 
a disk, the shape of the line allows us to constrain the region 
where the emission is produced if the inclination is known,
or to constrain the inclination of the disk if the
region where the emission is produced is given.

Implementing the two-layer Chiang and Goldreich (1997) 
disk model code CG plus (Dullemond et al. 2001),  
we modeled the disk around LkH$\alpha$ 264.
As inputs for the model we used,
a disk without a puffed-up inner rim with
an inner truncation radius at $T=$ 3000 K,
a disk size of 250 AU,
a mass\footnote{
Note that the disk mass employed here is deduced from mm dust continuum emission,
therefore, it is an estimate of the mass of the cold outer disk. Since most of the 85 M$_J$ are  located beyond a few tens of AU from the star and the total disk mass depends on the (unknown) value of the dust mass absorption coefficient and the gas-to-dust mass ratio, 
our parameterization of the disk would be a lower limit to the possible disk mass. Nevertheless, the uncertainty in the total disk mass has a small influence in the line fitting, because 
the contribution of the inner disk to the total
mass is minor and 
the H$_2$ emission is produced in the optically thin upper layer of the disk.} of 85 M$_J$,
a density power law factor of -1.5
and a luminosity of 0.53 log($L/L_{\odot}$) for LkH$\alpha$ 264.
The luminosity was determined from the 
spectral type K5V ($T_{eff}=4350$ K),
using a distance of 300 pc, 
a $V$ magnitude of 12 mag, an extinction A$_V$ = 0.5 mag (Itoh et al. 2003) 
and a bolometric correction of -0.72 (Kenyon and Hartmann, 1995).
We found that the regions of the disk with a surface layer at 
$T_s < 1500$ K are located at $R > $ 0.1 AU.

Prescribing a mass of 0.8 M$_{\odot}$ for LkH$\alpha$ 264
(found by its location in the HR diagram
employing the evolutionary tracks of Palla and Stahler 1993),
an H$_2$ emitting region from 0.1 to 10 AU, 
and assuming that the intensity $I$ of the H$_2$ line
is described by a power law according to the radius $I_{\rm H_2}(R)= I_0~R^{\alpha}$ 
($\alpha$ being a negative number)
we calculated the expected line profile produced by the inclined disk.
We proceed as follows. Suppose there is a parcel of gas  
situated at a radius $R$ of width d$R$ and
angular size d$\theta$ that emits a line intensity $I_{\rm H_2}(R)$ with a profile $\phi_\nu$, 
in a disk inclined at an angle $i$ surrounding
a star of mass $M_{\star}$ at a distance $D$ from the Earth.
The doppler shift $\Delta\nu$ of a line at frequency $\nu_0$ emitted by the parcel of
fluid is given by
\begin{equation}
{\Delta\nu}=\frac{\nu_{0}~sin(\theta)sin(i)}{c}\sqrt{\frac{GM_{\star}}{R}},
\end{equation}  
with $c$ being the speed of light.
Assuming that the line profile $\phi_\nu$ is Gaussian,
the doppler shifted line profile $\phi_{\nu {\rm~shifted}}$ 
emitted by the parcel of fluid is
\begin{equation}
\phi_{\nu~{\rm shifted}}=\frac{1}{{\rm \sigma_{\nu}} \sqrt{\pi}}e^{-\left(\frac{\nu - \nu_0+\Delta\nu}{\rm \sigma_{\nu}}\right)^2},
\end{equation}
where $\sigma_{\nu}= (\nu_0/c)*{\rm FWHM}/\left(2\sqrt{\rm ln2}\right)$.
The FWHM (in km s$^{-1}$) in strict terms is the thermal broadening of the line,
however, for the purpose of calculating of the observed profile by the instrument,
we assumed the FWHM to be the resolution of the spectrograph (in our case 6.6 km s$^{-1}$).
The flux emitted by the parcel of fluid is therefore
\begin{equation}
dF_{\rm H_2}(R,\theta)=I_{\rm H_2}(R)~\phi_{\nu~{\rm shifted}}~\frac{R\,dR\,d\theta}{D^2},
\end{equation}
and the total emitted line flux is the sum of the contributions of all the fluid parcels
\begin{equation}
F_{\nu~{\rm H_2}}=\int^{R_{max}}_{R_{min}}\int^{2\pi}_{0}dF_{\rm H_2}.
\end{equation} 
Here, $R_{min}$ and $R_{max}$ are the inner and outer radius of the region responsible for the emission.
For our calculation we assumed that $R_{min}=0.1$~AU, $R_{max}=10$~AU, 
$I_0=1$ and $D=300$~pc. The results are weakly dependent on the selection of a larger $R_{max}$
since the intensity $I$ decreases rapidly as a function of radius. 
$R_{min}$ is selected as 0.1 because it is at this radius at which the temperature starts to be cooler than 1500 K in the disk
surface of LkH$\alpha$ 264.
The resulting synthetic line profile was scaled in such a way that the peak flux
of the synthetic line is equal to the peak flux of the line (minus continuum) observed.
The $\alpha$ exponent of the intensity as a function of radius was assumed to 
be equal to -3, -2 and -1 for each set of models. 
  
With the inclination $i$ being the only free parameter,
we manually changed its value for each value of $\alpha$ until we 
found a good fit for the line. If the inclination selected was too large, a double peaked profile
was obtained, if it was too small, the velocity wings and the width of the
line obtained were too narrow. Thus, only a small interval of inclinations fit the line profile
for each value of $\alpha$.  
We found that for reproducing the observed line profile,
the inclination of the disk should be close to face-on, from 
8$^{\rm o}$ to 35$^{\rm o}$ for $\alpha$ power law exponents ranging 
from -3 to -1 respectively.
In Figure 2 we present the best fit found: an inclination of 
20$^{\rm o}$ and $\alpha=-2$.
The close to face-on inclination derived from our CRIRES data
is consistent with the polarization measurements of Bastien (1982), 
who found that the polarization degree of LkH$\alpha$ 264 in the optical is fairly 
low ($<1\%$), which is consistent with a small inclination (p is zero for i=0).
If the $\upsilon = 1-0$ H$_2$ S(1) line intensity decreases with an exponent $\alpha=-2$ as a function of radius,
then 50\% of the line flux is produced within 0.1 AU and 1 AU of the LkH$\alpha$ 264 disk,
40\% of the line flux is emitted within 1 and 7 AU and the remainder of the flux (10\%) 
is emitted at larger radii. 

\begin{figure}
\centering
\includegraphics[angle=0,width=0.5\textwidth]{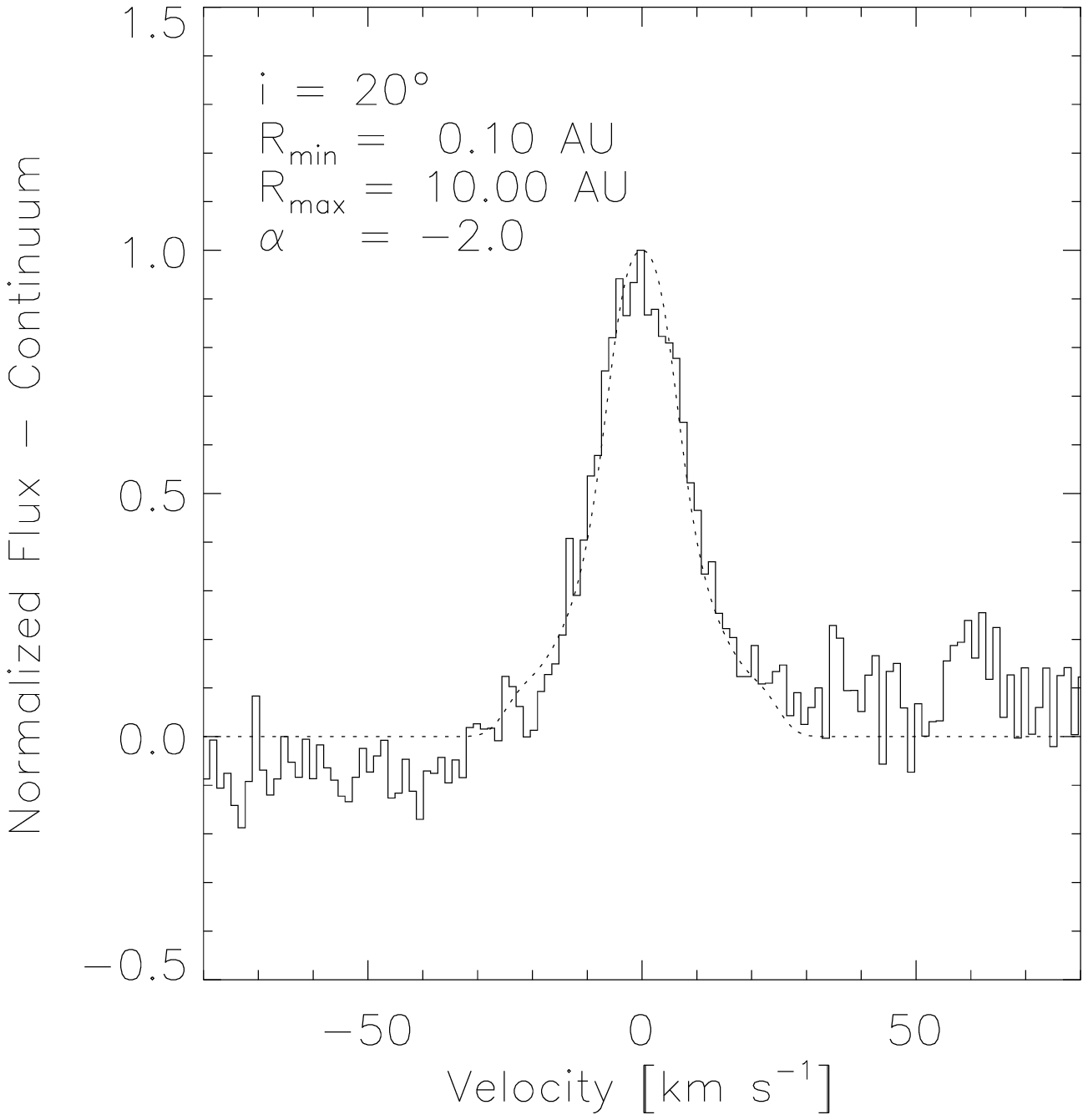}
    \caption{Best model fit for the H$_2~\upsilon~1-0$ S(1) line detected in
            LkH$\alpha$ 264 assuming that the emission originates in a circumstellar disk. 
            R$_{\rm min}$ and R$_{\rm max}$ are the inner and outer 
            radius of the emitting region. $\alpha$ is the power
            law exponent of the intensity $I(R) \propto R^{\alpha}$.}
\end{figure}
\begin{table*}
\caption{Physical properties of classical T Tauri stars in which a search for H$_2$ $\upsilon = 1-0$ S(1) emission was performed.}             
\label{table:1}      
\centering                          
\begin{tabular}{l c c c c c c c c c c c c c c c c }        
\hline\hline                 
\addlinespace[2pt]
      &   &   $\dot{M}$& EW H${\alpha}$ & A$_{V}$ & $(U-V)_{\rm obs}$ & $(U-V)_{\rm dered}$ & $(U-V)_{ex}$ 
      & log\,$L_{\rm X}$ & $M_{disk}^{\dag}$ &  \\
Star            & Sp.T. & [$\times~10^{-7}$ M$_{\odot}$ yr$^{-1}$] & [\AA]  & [mag]        & [mag]       & [mag]         & [mag] 
	  & [erg s$^{-1}$]   & [M$_J$] & Ref. H$_2$ \\
\hline
\addlinespace[3pt]                                   
\multicolumn{11}{c}{{\it Detections}}\\
\addlinespace[3pt]                                   
\hline
\addlinespace[3pt]                                   
LkH$\alpha$ 264  & K5 Ve  & \,\,\,\,\,1 - 10 $^{af}$ & 85 $^{h}$ & 0.52 & 0.37 & 0.08  & -2.10  & 29.7  & 85$^q$ & 1,2 \\
TW Hya           & K7 Ve  & \,\,\,\,\,\,0.005 $^{ag}$ & 86 $^{h}$ & ~~0.18$^a$ &  0.86 & ~~0.8$^a$   & ~-1.8$^a$  & 30.3  & 1.4$^r$ &3\\
GG Tau Aa        & K7 Ve  & \,\,\,\,\,\,0.175 $^{\dag\,\dag}$& 54 $^{h}$& ~~3.20$^a$ &  2.73 & ~~1.0$^a$ & ~-1.6$^a$ & 29.4 & 290$^p$ & 3,6 \\  
GG Tau Ab        & M0.5 Ve& \,\,\,\,\,\,0.175 $^{\dag\,\dag}$& 54 $^{h}$& ~~0.72$^a$ &  1.42 & ~~1.0$^a$ & ~-1.5$^a$ & 29.4 & 290$^p$ & 3,6 \\
LkCa 15          & K5 Ve  & \,\,\,\,\,\,0.015 $^{ah}$& 13 $^{h}$   & 0.64 $^{h}$&  ~~1.98$^a$ & ~~1.6$^a$ & ~-0.5$^a$ & $<$29.6$^e$ & 10$^s$ & 3,6 \\
AA Tau$^b$       & K7-M0 Ve  & \,\,\,\,\,\,0.033 $^{\dag\,\dag}$& 37 $^{b}$  & 0.93 & ~~0.9$^c$ & 0.4 & ~-2.2$^d$  & ~~29.6$^e$ &21$^p$ & 6 \\
CW Tau$^b$       & K3 Ve     & \,\,\,\,\,\,0.016 $^{ai}$ & 135 $^{b}$ & 2.34 & ~~1.46$^c$ & 0.15 & -1.65  & ~~30.5$^l$ & $<$15$^p$  & 6 \\
UY Aur$^b$       & K7 Ve     & \,\,\,\,\,\,0.656 $^{\dag\,\dag}$& 73 $^{b}$ & 1.05 & ~~0.92$^f$ & 0.33 & -2.18  & $<$29.4$^e$ & 0.9$^t$    & 6 \\
GM Aur$^b$       & K7-M0 Ve  & \,\,\,\,\,\,0.096 $^{\dag\,\dag}$& 97 $^{b}$ & 0.14 & ~~1.5$^g$ & 1.4& ~-1.2$^d$  & $<$29.7$^e$ & 60$^p$ & 6 \\
CS Cha           & K5 Ve     & \,\,\,\,\,\,\,\,\,\,1.6 $^{aj}$ & 13 $^{h}$ & 0.06$^i$ & ~~1.72$^i$ & 1.68 & ~0.49  & ~~30.2$^v$ & 21$^w$ & 7 \\
ECHA J0843.3-7905 & M3.2 Ve$^x$   & \,\,\,\,\,0.010 $^{ak}$ &  111 $^z$  & $<$0.1$^{ac}$  & $^{ad}$& ... & ... & $<$28.5$^{aa}$ &$^{ae}$ & 8 \\
\hline
\addlinespace[3pt]                                   
\multicolumn{11}{c}{{\it Non-Detections}}\\
\addlinespace[3pt]                                   
\hline
CD -33$^{\rm o}$7795  & M1.5Ve    & \,\,\, $<0.1$ $^{al}$ & 15 $^m$  & 0.07$^n$   & 2.39$^n$   & 2.35 & -0.33 & 30.6    & $<$0.1$^u$ & 5  \\   
IP Tau                & M0 Ve     & \,\,\,\,\,\,0.008 $^{\dag\,\dag}$ & 11 $^{h}$      & ...$^y$    & 2.04$^y$   &  2.04$^y$ & 0.63 & 29.5    & 5$^t$ & 3 \\  
IQ Tau$^b$			   & M0.5 Ve   & \,\,\,\,\,\,\,\,\,\,0.18 $^{aj}$& 8 $^{b}$      & 0.77       & 0.93       & 0.49 & -2.17 & $<$29.5~ & 40$^p$ &3 \\
V836 Tau$^{ha}$          & K7 V      & \,\,\,$<$ 0.001 $^{\dag\,\dag}$& 9 $^h$   & 0.71$^i$   & 2.71$^i$   & 2.31 & -0.21 & 29.8    & 40$^t$ &3\\
RECX 5                & M4.0 Ve$^x$   & \,\,\,\,\,0.0005 $^{ak}$ &  9 $^z$    & $<$0.1$^{ac}$  & $^{ad}$& ... & ... & ~~29.0$^{ab}$ &$^{ae}$ & 8 \\
RECX 9                & M4.5 Ve$^x$   & \,\,\,\,\,0.0004 $^{ak}$&  12 $^z$   & $<$0.1$^{ac}$  & $^{ad}$& ... & ...& ~~28.4$^{ab}$ &$^{ae}$ & 8 \\
\hline
\hline
\addlinespace[5pt]
\multicolumn{11}{l}{References : (1) Carmona et al. 2007 
(this work); (2) Itoh et al. (2003); (3) Bary et al. (2003); (4) Bary et al. (2002); }\\
\multicolumn{11}{l}{(5) Weintraub et al. (2000); (6) Shukla et al. (2003); (7) Weintraub et al. (2005). (8) Ramsay
Howat \& Greaves (2007)}\\
\multicolumn{11}{l}{Notes: $^{\dag} M_{disk}$ refers to the total mass in the disk deduced from 
mm dust continuum emission assuming a gas to dust ratio of 100.}\\
\multicolumn{11}{l}{$^a$ Average value from Table 5 of Bary et al. (2003);
\,$^b$ Spectral Type, H$\alpha$ EW and A$_V$ from Cohen \& Kuhi (1979);}\\
\multicolumn{11}{l}{$^c$ Varsavsky (1960); $^d$ Average between the excess of the two spectral types; $^e$ Neuh\"auser et al. (1995); $^f$ Mendoza (1966);}\\
\multicolumn{11}{l}{$^g$ Bastian \& Mundt (1979); 
$^h$ H$\alpha$ EW by Herbig \& Bell (1988); $^{ha}$ Given the H$\alpha$ EW  we classified
the source as CTTS ;}\\
\multicolumn{11}{l}{\,$^i$ Herbig \& Bell (1988); $^j$ Quadruple system (Soderblom et al. 1998); $^k$ Mermilliod (1986);
\,$^l$ XEST data by  G\"udel et al. (2007);}\\
\multicolumn{11}{l}{$^m$ Craig et al. (1997); $^n$ Gregorio-Hetem et al.(1992); $^p$ Beckwith et al. (1990); $^q$ Itoh et al. (2003b);  }\\
\multicolumn{11}{l}{$^r$ Weinberger et al. (2002); $^s$ Qi et al. (2003);
\,$^t$ Osterloh \& Beckwith (1995);}\\
\multicolumn{11}{l}{$^u$ Jayawardhana et al. (1999), Weinberger et al. (2004) and Uchida et al. (2004) do not find evidence for infrared 
excess in the}\\
\multicolumn{11}{l}{source. Here we adopt the lower limit on the disk's masses of  Osterloh \& Beckwith (1995) as
upper limit for the disk's mass.}\\
\multicolumn{11}{l}{$^v$ Costa et al. (2000); $^w$ Lommen et al. (2007);
\,$^y$ ROTOR data (Grankin et al. 2007) gives an E$(B-V)$ of -0.74 
which is more}\\
\multicolumn{11}{l}{than a magnitude off from that of Herbig \& Bell (1988).
Maybe there is 
some long-term evolution of this system that changes its}\\
\multicolumn{11}{l}{colors on a long time scale.
We worked with the $U-V$ ROTOR color but did not correct for redenning when 
calculating the $U-V$ excess;}\\
\multicolumn{11}{l}{$^x$ Luhman \& Steeghs (2004); $^z$ Jayawardhana et al. (2006); $^{aa}$ Lawson et al. (2002);
$^{ab}$ Mamajek et al. (1999); $^{ac}$ Lyo et al. (2004);}\\
\multicolumn{11}{l}{$^{ad}$ No $U$ photometry published; $^{ae}$ No 1.3 mm continuum flux published. $^{af}$ Accretion rate from Gameiro et al. (2002);}\\
\multicolumn{11}{l}{$^{ag}$ Muzerolle  et al. (2000); $^{ah}$ Akeson et al. (2005); $^{ai}$ Valenti et al. (1993);
$^{aj}$ Johns-Krull et al. (2000); $^{ak}$ Lawson et al. (2004);}\\
\multicolumn{11}{l}{$^{al}$ Mohanty et al. 2003
detected H$\alpha$ in emission of 10\% width $\sim$ 270 km s$^{-1}$. 
Since no broad OI (8446 \AA) or Ca II 
(8446) emission was }\\
\multicolumn{11}{l}{reported by those authors, $\dot{M}$ should be lower 
than 10$^{-8}\,M_{\odot}$ yr$^{-1}$ (Jayawardhana et al. 2003); $^{\dag\,\dag}$ Gullbring et al. (1998).}
\end{tabular}
\end{table*}

\begin{figure*}
\centering
\includegraphics[angle=0,width=0.7\textwidth]{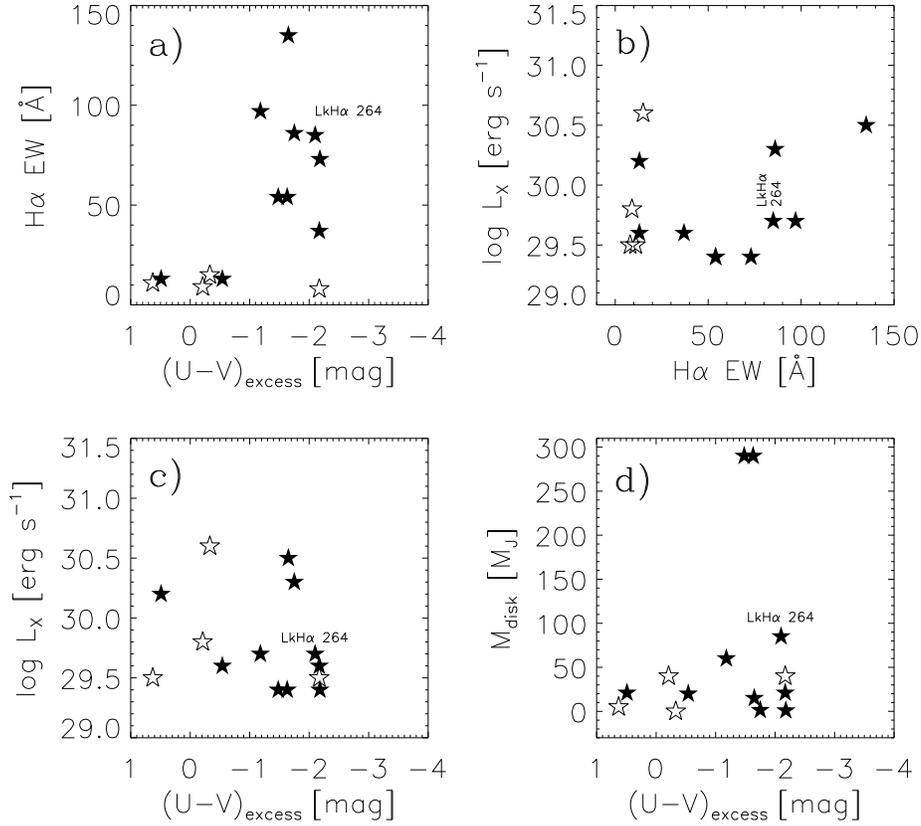}
    \caption{Physical properties of classical T Tauri stars in which a search for the H$_2$ $\upsilon =1-0$ S(1) 
    line was performed. Filled stars represent detections, non-filled stars
    represent non-detections. (a) H$\alpha$ EW versus $(U-V)_{\rm excess}$. 
    (b) log$L_X$ [erg s$^{-1}$] versus H$\alpha$ EW.
    (c) log$L_X$ [erg s$^{-1}$] versus $(U-V)_{\rm excess}$.
    (d) $M_{disk}$ versus $(U-V)_{\rm excess}$.               }
\end{figure*}
\subsection{H$_2$ NIR Ro-vibrational Emission in LkH$\alpha$ 264 and Other T Tauri Disks}

The mass determination of H$_2$ gas in the inner disk 
is crucial for  constraining the properties of the gas 
in the terrestrial planet forming region.
However, the detections of ro-vibrational H$_2$ emission from disks are relatively 
scarce compared to the large number of pre-main-sequence stars with gas-rich disks
that are known.
So far the $\upsilon =$1-0 S(1) line has been detected in 
few classical T Tauri stars (CTTS): TW Hya, GG Tau A, LkCa 15 (Weintraub et al. 2000, Bary et al. 2002, 2003),
AA Tau, CW Tau, UY Aur, GM Tau (Shukla et al. 2003), CS Cha (Weintraub et al. 2005), ECHAJ0843.3-7905 (Ramsay Howat \& Greaves 2007) and LkH$\alpha$ 264 (Itoh et al. 2003, this paper),
and in four weak-line T Tauri stars (WTTS): DoAr 21 (Bary et al. 2003), V773 Tau (Shukla et al. 2003),
Sz33 and Sz 41 (Weintraub et al. 2005).
Our CRIRES observations show, for the first time, the simultaneous detection of the
$\upsilon = 1-0$ S(1) and $\upsilon = 1-0$ S(0) H$_2$ emission from a protoplanetary disk.
Since the detections are not very numerous, it would be useful to know if the T Tauri
stars with detected H$_2$ near-infrared ro-vibrational emission are peculiar objects.

In the case of H$_2$ emission detected in WTTS, 
DoAr 21 and V773 Tau are among the brightest X-ray WTTS (see Table 1 and 4 of Bary et al. 2003 and for
V773 Tau the XEST data of G\"udel et al. 2007)\footnote{log $L_X$ of V773 Tau is 31.0 erg s$^{-1}$. The star is 
a quadruple system (Duch\^eme et al. 2003). The K-type binary is expected to widely dominate in 
X-rays (Audard, private communication).}.

In the case of CTTS, such a correlation is not apparent.  
In Table 4, we summarize some important physical properties of the CTTS
in which a search was done for H$_2$ $\upsilon = 1-0$ S(1) emission\footnote{At the time of writing,
non-detections have been only reported in Weintraub et al. (2000) and Bary et al. (2003). 
In the cases of the observations by Shukla et al. (2003) and Weintraub et al. (2005) only the 
names of the stars in which H$_2$ was detected are published.}.
We list properties related to the accretion process such as the accretion rate, 
H$\alpha$ emission EW and the $U-V$ excess.
In addition, we present the X-ray luminosity and the disk's mass deduced from mm
dust continuum emission reported in the literature. 
For the calculation of the $U-V$ excess, 
we first determined the $U-V$ dereddened color employing the 
visual extinction $A_{V}$ assuming an interstellar medium extinction law ($A_U = 1.56 A_V$).
Thereafter we subtracted from the $U-V$ dereddened color the $U-V$ color intrinsic to the spectral type 
of the source by Johnson (1966). 
In the case of multiple spectral types for a source, we selected their average value.

With the intention of unveiling empirical correlations 
between the physical properties of the sources 
and the detectability of the H$_2$ $\upsilon =1-0$ S(1) line,
employing the data collected in Table 4, 
we created a series of plots relating the physical properties  
of the sources (see Figure 3).

Two possible mechanisms of excitation have been proposed
as responsible for the H$_2$ emission in disks: X-ray and UV excitation.
We observe in Figure 3 (panels b and c) that in the case of the CTTS there is no clear correlation
between the X-ray luminosity and the detectability of the H$_2$ line.
We have sources with faint X-ray luminosity and H$_2$ detections (e.g. CW Tau) and
sources with relatively high X-ray luminosity but without H$_2$ detections (e.g. CD -33$^{\rm o}$7795)\footnote{
Note that in the case of CD -33$^{\rm o}$7795, it could be argued that there is no detection because there
is no disk: this source does not show infrared excess (Jayawardhana et al. 1999; Weinberger et al. 2004; 
Uchida et al. 2004).
However the source does exhibit H$\alpha$ in emission Craig et al. (1997) and $U-V$ excess.}.
In addition, in several sources with X-ray luminosities smaller than that of V836 Tau (a non-detection) 
the H$_2$ line has been detected.
We conclude that X-ray excitation could play a role in the heating of the gas, 
but that in the case of CTTS studied so far, it seems to not be the dominant factor in the excitation of NIR H$_2$ emission.

The second source for the excitation of H$_2$ emission is UV photons. 
UV photons are produced in large quantities
during the accretion process. The $U-V$ excess and the H$\alpha$ emission are considered
standard signatures of accretion in T Tauri stars. 
In Figure 3a we show the H$_{\alpha}$ EW vs the $U-V$ excess. 
We observe that the higher the $U-V$ excess and the stronger the H$\alpha$ line are, 
the higher the number of sources with H$_2$ detections.
The non-detections are situated in the area of small H$_{\alpha}$ EW and low $U-V$ excess.
{\it This result suggests that the higher the accretion rate in the systems is, 
the higher the probability of  exhibiting the $\upsilon = 1-0$ S(1) H$_2$ line}.
For example the only object exhibiting  H$_2$ 2.12 $\mu$m emission in the
$\eta$ Chamaeleontis cluster is ECHA J0843.3-7905, a source  with strong H$\alpha$ emission
and a comparatively high accretion rate (10$^{-9} M_{\odot}$ yr$^{-1}$, Lawson et al. 2004).  
We should note that there are detections of the H$_2$ line 
in two objects (CS Cha and LkCa 15) that are located in the region 
of the H$\alpha$ vs $U-V$ excess diagram where three non-detections are situated.
In the case of CS Cha, the cause of the emission is probably the high X-ray luminosity.
In the case of LkCa 15 no X-rays have been detected. 
But, LkCa 15 is an edge-on disk source. The lack of X-ray luminosity and of a large U-V excess for this star may be related to the disk geometry.
It is also interesting to realize that there is a non-detection in a source (IQ Tau) that 
has a strong $U-V$ excess. However, IQ Tau exhibits a very small H$\alpha$ EW. 

Our CRIRES target, LkH$\alpha$ 264, is  one of the sources with the strongest $U-V$ excess in the sample.
With respect to other physical characteristics (H$_{\alpha}$ EW, disk mass and X-ray luminosity),
LkH$\alpha$ 264 is a "normal" source. 
Therefore, it is likely that in LkH$\alpha$ 264 UV photons are mainly responsible for the H$_2$ emission.
This conclusion is supported independently by the measured 1-0 S(0)/1-0 S(1) and 2-1 S(1)/1-0 S(1) line ratios 
as previously discussed.

Concerning the disk mass and the detectability of the H$_2$,
there is no apparent correlation between them. H$_2$ detections and non-detections are present in  
the disk mass range from 1 to 40 M$_J$. 
In summary,  
LkH$\alpha$ 264 and the CTTSs in which H$_2$ emission has been detected share
typical physical properties of classical T Tauri stars. 
Therefore, in the near future, we expect to see more detections of the H$_2$ near-infrared lines 
to come out of high-resolution spectrographs on a routine basis.

\subsection{49 Cet Disk}
49 Cet is a young isolated star with age between ~10 Myr and ~100 Myr.
Its position on the colour-magnitude diagram with field stars and a few
well-studied clusters indicates that it is roughly
intermediate in age between IC 2391 and the Pleiades.  
It is ~40 pc below the Galactic plane, within the Local Bubble, 
and its velocity is inconsistent
with being near Sco-Cen (or its cloud complexes) or Taurus in the recent
past.
While its velocity would be very discrepant for a
very young system ($<$20-30 Myr), 
its UVW is in the ballpark of some
known $\sim$30-80 Myr systems (e.g., IC 2391, NGC 2451A). 
However, it's velocity is not near that of the Gould Belt ($\lesssim$ 30 Myr groups
within $\sim$0.5 kpc; Mamajek, private communication).  

Optical spectroscopy of 49 Cet basically shows
the typical spectrum of an A-type main sequence star.
49 Cet does not exhibit H$\alpha$ in emission
and does not present UV excess in its Spectral Energy Distribution (SED).
49 Cet, therefore, is very likely not accreting.
In the JHK bands the colors of 49 Cet do not differ significantly
with respect to the JHK colors of an A1V star.
However, 
in the  mid- and far-infrared (i.e., 25, 60 and 100 $\mu$m)
49 Cet exhibits emission in excess of photospheric levels,
thereby revealing the existence of a circumstellar disk. 
Recent analysis of sub-arcsec mid-infrared imaging of 49 Cet by Wahhaj et al. (2007)
suggests that the bulk of the mid-infrared emission comes from very small grains
($a\sim$ 0.1  $\mu$m) confined between 30 and 60 AU from the star,
and that most of the non-photospheric flux is radiated by an outer disk
of large grains ($a\sim$ 15 $\mu$m) of inner radius $\sim 60$ AU and outer radius 900 AU.
In their analysis Wahhaj et al. (2007) conclude that the most likely scenario 
is that the inner 20 AU is strongly depleted of dust.
  
Zuckerman and Song (2004) proposed an age of 20 Myr for 49 Cet,
an age in which the gaseous disk is expected to have already dissipated.
However, Zuckerman et al. (1995) and  Dent et al. (2005) 
observed CO $J=2-1$ and $J=3-2$  emission towards
49 Cet, thereby revealing the existence of a reservoir of cold gas.
Dent et al. (2005) modeled the double peaked CO $J=3-2$ emission and proposed 
that the line is emitted from a compact disk of outer radius $\sim$17 AU inclined at 16$^{\rm o}$ 
or a disk of outer radius $\sim$50 AU but inclined at 35$^{\rm o}$. 
Thi et al. (2001) based on ISO measurements 
claimed the detection of the pure rotational 0-0 S(0) H$_2$ emission
at 28 $\mu$m in 49 Cet, but recent more sensitive {\it Spitzer} IRS observations by Chen et al. (2006) did not confirm the detection of the line.

The detection of CO emission in the sub-mm and the apparent 
existence of a dust gap in the interior of the 49 Cet disk
pose the question whether gas still exists within the inner disk (R $<$ 20 AU).
The upper limit on the flux of the H$_2$ ro-vibrational lines in 49 Cet
derived from our CRIRES data
set stringent constraints on the amount of hot gas in the 
inner disk of 49 Cet (R $<$ 1 AU):
49 Cet has less than a tenth of lunar mass of gas at T$\sim$1500 K.
Our observations give additional support to the hypothesis that 
the disk of 49 Cet has an inner hole.
The lack of H$_\alpha$ in emission and the non-detection 
of pure rotational and 
ro-vibrational emission of warm and hot H$_2$ in 49 Cet
indicate that 49 Cet may have an inner hole in gas as well\footnote{The lack of H$_2$ 
emission means, either the dissipation of H$_2$ or lack of excitation mechanism. We favor
the dissipation of H$_2$ because the H$_2$ NIR lines are sensitive to very small amounts
of gas. Only a few moon masses of optically thin hot ($T\sim 1500$ K) 
H$_2$ will provide detectable line fluxes.  
In addition, this interpretation is consistent with the 
lack of H$_\alpha$ in emission and the non-detection 
of pure rotational emission of warm H$_2$ in 49 Cet.}. 
This result supports the idea that gas and dust are 
dissipated on the same time scale in the inner disk
(Sicilia-Aguilar et al. 2006), 
and is strongly suggestive that the disk disappears inside-out.

One interesting question to address is the possible mechanism of disk dissipation.
It has been suggested (e.g., Alexander et al. 2006) that inside-out photoevaporation 
occurs very rapidly in a time scale of a few 10$^5$ years once the phenomenon is 
triggered after a disc life time of few million years. 
A challenge to this scenario is the presence of CO in the outer disk.
Once the photoevaporation starts in the inner part of the disk,
the entire gaseous disk should dissipate 
in a very short time frame as well (Alexander et al. 2006). 
Therefore, one puzzling aspect in this scenario is the reason 
why the outer gas remains in the system.
An alternative scenario is to assume the presence of a sub-stellar companion
to explain the lack of gas in the inner disk.
This explanation has the advantage that the presence of a gas rich outer disk is
a natural part of the planet formation process. 
Giant planets are thought to form in the inner 20 AU of the disk and the 
outer disk disappears later
once the planets have been formed.
We note that indications for the presence of a planet in a disk, 
as required for this scenario,
have recently been found in a precursor of a 49-Cet type star, 
the Herbig Ae/Be star HD 100456 (Acke \& van den Ancker 2006).
The existence of low mass companion(s) as an explanation for the lack of
gas and dust in the inner disk of 49 Cet is a suggestive idea that, 
given the relative closeness of the target ($d\sim$ 61 pc), 
it will be possible to test. 
For example, Apai et al. (2007) made a sensitive adaptive optics search 
for close companions to a sample of 8 nearby cold debris disks ($d=$ 20 - 70 pc) 
and found no evidence for companions of masses higher
than 3 - 7  M$_{J}$ and higher at separations larger than 15 AU.
Future high-contrast imaging facilities such as SPHERE at ESO-VLT, will 
allow to search for lower mass companions at closer separations.
 
\section{Conclusions}
We observed the classical T Tauri star LkH$\alpha$ 264 and the debris
disk 49 Cet and searched for ro-vibrational $\upsilon =1-0$ S(1) H$_2$ emission at 2.1218 $\mu$m,
$\upsilon =1-0$ S(0) H$_2$ emission at 2.2233 $\mu$m,
and $\upsilon = 2-1$ S(1) H$_2$ emission at 2.2477 $\mu$m,
using CRIRES ($R \sim 6.6$~km s$^{-1}$) at ESO-VLT.
We confirmed the detection of  the $\upsilon =1-0$ S(1) H$_2$ line in LkH$\alpha$ 264
at the rest velocity of the star.
The line has a flux of 3.0 $\times 10^{-15}$ ergs cm$^{-2}$ s$^{-1}$,
and a FWHM of 20.6 km s$^{-1}$.
In addition, the enhanced sensitivity of CRIRES allowed the observation
of the previously undetected $\upsilon =1-0$ S(0) H$_2$ line in LkH$\alpha$ 264.
The line has a flux of 1.0 $\times 10^{-15}$ ergs cm$^{-2}$ s$^{-1}$,
and a FWHM 19.8 km s$^{-1}$.
An upper limit of 5.3 $\times 10^{-16}$ ergs s$^{-1}$ cm$^{-2}$ 
was derived for the $\upsilon = 2-1$ S(1) H$_2$ line flux in LkH$\alpha$ 264.
The very similar FWHM of the two H$_2$ lines detected suggests that the emitting 
gas is located in similar regions in the disk.
Both lines are spatially unresolved.
The measured mean PSF's FWHM ($\approx$ 0.36") in the H$_2$ 1-0 S(1) spectrum
indicates that the H$_2$ emitting region
is located in the inner 50 AU of the disk assuming a distance of 
300 pc for LkH$\alpha$ 264.
The measured 1-0 S(0)/1-0 S(1) (0.33 $\pm$ 0.1)
and the 2-1 S(1)/1-0 S(1) ($<$0.2) line ratios
in LkH$\alpha$ 264 indicate 
that the H$_2$ emitting gas 
is at a temperature lower than 1500 K and that the H$_2$ is most likely 
thermally excited by UV photons. 
The measured line ratios suggest that X-ray excitation plays only a minor role
in the heating of the emitting H$_2$ in LkH$\alpha$ 264.
The flux of the $\upsilon =1-0$ S(1) H$_2$ line in LkH$\alpha$ 264 implies that
there are a few lunar masses of hot H$_2$ gas in the inner disk of LkH$\alpha$ 264.
The $\upsilon =1-0$ S(1) H$_2$ line in LkH$\alpha$ 264 is single peaked.
Modeling of the $\upsilon =1-0$ S(1) line shape indicates that the disk is close to face-on
($i<35^{\rm\,o}$).
The best model fit suggests that the disk of LkH$\alpha$ 264 is inclined 20$^{\rm\,o}$ 
for a H$_2$ emitting region extending from 0.1 to 10 AU with a power law relation
of the intensity as a function of radius with exponent $\alpha=-2$.
If the $\upsilon = 1-0$ H$_2$ S(1) line intensity decreases with an exponent $\alpha=-2$ as a function of radius,
then 50\% of the line flux is produced within 0.1 AU and 1 AU of the LkH$\alpha$ 264 disk,
40\% of the line flux is emitted within 1 and 7 AU and the rest of the flux at larger radii.

A comparative analysis of the physical properties of classical T Tauri stars
in which the H$_2$ $\upsilon =1-0$ S(1) line has been detected versus non-detected
shows that there is a higher chance of observing the H$_2$ near-infrared 
lines in CTTS  with a
high $U-V$ excess and a strong H$\alpha$ line. 
This result suggests that there is a higher probability of detecting the
H$_2$ $\upsilon =1-0$ S(1) line in systems with high accretion.
In contrast to weak-lined T Tauri stars, 
there is no apparent correlation between the X-ray luminosity and
the detectability of the H$_2$ $\upsilon =1-0$ S(1) line in classical T Tauri stars.
Taken as a group, 
LkH$\alpha$ 264 and the CTTS in which the H$_2$ emission has been detected exhibit
typical properties of classical T Tauri stars. 
Therefore, we expect NIR ro-vibrational H$_2$ lines from T Tauri disks 
to be detected on a routine basis in the near future.
   
The non-detection of any of the three H$_2$ lines in 49 Cet  
puts stringent constraints on the amount of hot gas in the inner disk.
From the upper limit for the flux of the $\upsilon =1-0$ S(1) H$_2$ line we deduced that
less than a tenth of lunar-mass of gas at $T\sim 1500$ K is present in the inner 1 AU 
of the disk surrounding 49 Cet.
The lack of H$_2$ ro-vibrational  emission in the  spectra of 49 Cet, 
combined with non detection of pure rotational lines of H$_2$ (Chen et al. 2006)
and the absence of H$\alpha$ emission suggest that the gas in the 
inner disk of 49 Cet has dissipated.
These results together with the previous detection of $^{12}$CO emission at sub-mm wavelengths 
(Zuckerman et al. 1995; Dent et al. 2005) point out that the disk of 49 Cet should have a large inner hole,
and it is strongly suggestive of theoretical scenarios in which the disk disappears inside-out.
We favor inner disk dissipation by inside-out photoevaporation, or
the presence of an unseen low-mass companion(s) as most likely explanations
for the lack of warm gas in the inner disk of 49 Cet.

\begin{acknowledgements}
This research has made use of the SIMBAD database
operated at CDS, Strasbourg, France.
A.C. would like to thank R. Mund for helpful discussions concerning outflows
in T Tauri stars, M. Audard for kindly providing XEST X-ray luminosities of several sources,
G. van der Plas for calculations of the rotational broadening of absorption lines,
and C. Fallscheer for comments to the manuscript. 
Special thanks to the CRIRES science-verification team for executing
the observations in Paranal and for their support in the data-reduction process.
\end{acknowledgements}

\end{document}